\input{aipcheck}

\documentclass[
    ,final            
  ]
  {aipproc}

\layoutstyle{8x11double}

\begin{document}

\title{Fermi Acceleration at Relativistic Shocks}


\author{Guy Pelletier}{
  address={Laboratoire d'Astrophysique, Universit\'e Joseph Fourier, B.P. 53, 38041 Grenoble, France}
}

\author{Martin Lemoine}{
  address={Institut d'Astrophysique de Paris, 98 bis, bd Arago, 75014 Paris, France}
}

\author{Alexandre Marcowith}{
 address={Laboratoire d'Astroparticules et Physique Th\'eorique, Pl E. Bataillon, 34094 Montpellier, France}
}

\begin{abstract}
After a successful development of theoretical and numerical works on Fermi acceleration at relativistic shocks, some difficulties recently raised with the scattering issue, a crucial aspect of the process. Most pioneering works were developed assuming the scattering off magnetic fluctuations as given.  Even in that case, when a mean field is considered, its orientation is mostly perpendicular to the shock normal in the front frame, and this tends to quench the scattering process. Solving this difficulty leads to address the issue of the generation of very intense magnetic fluctuations at short wave lengths. The relativistic motion of the shock front let the cosmic rays to visit upstream during a very short time only, making this generation of magnetic fluctuations very challenging. Anyway there is some hope to solve the problem. Thanks to a recent work by Spitkovsky (2008) \cite{AS}, we know that the process works without any mean field and now we have to investigate up to which  intensity the mean field can be amplified for allowing Fermi process with appropriate fast instabilities.
In this presentation, the collisionless shock structure in relativistic regime is sketched, the scattering issue is presented, and the instabilities that can provide the expected magnetic field amplification are presented as well. Although there exists observational evidence that particles are accelerated in relativistic flows and are distributed according to a power law suggesting a Fermi process, the drastic conditions for Fermi process to work are not always clearly fulfilled.
 
 \end{abstract}

\maketitle


\section{Introduction}

Important progress were accomplished during this last decade about the acceleration of particles governed by Fermi process at relativistic shocks  \cite{BO}, \cite{GAC}, \cite{AGKG}, \cite{KGGA}, \cite{ED}. In these works a prescribed scattering process was introduced in the calculations and in the numerical works. In Lemoine \& Pelletier 03 \cite{LP} a scattering medium was simulated by introducing magnetic fluctuations with Fourier modes distributed according to a power law spectrum. The same results were obtained, in particular the spectrum index between $2.2$ and $2.3$; however with the assumption that the pitch angle only is involved in the scattering process; which was also assumed by the previous works mentioned. More detailed simulations by Niemiec, Ostrowsky \cite{NO}, then  Niemiec, Ostrowski and Pohl \cite{NOP}, and also by Lemoine, Revenu \cite{LR}, then Lemoine, Pelletier, Revenu \cite{LPR} revealed the difficulties. 
\begin{itemize}
  \item 
The mean field suffered by particles is essentially transverse to the flow in the front frame, if the angle $\theta_B$ of the mean field with respect to shock normal be larger than $1/\Gamma_s$, which is obviously the most frequent case.
\item
The downstream magnetic field is very compressed and weakens particle diffusion, therefore the universal spectrum is not obtained \cite{NO} \cite{NOP} \cite{LR}.
\item
An usual turbulence cascade spectrum from large scales to small scales cannot provide the expected scattering \cite{NO} \cite{LPR}. 
\end{itemize}
Finding circumstances favorable for Fermi process at a relativistic shock is very challenging.

Because it is useful for these issues, we first indicate what a collisionless shock structure is, especially in proton-electron plasma with a quasi perpendicular mean field. Then we present the scattering issue and the strong challenge for having Fermi cycles. Then we propose several fast instabilities that are good candidates for amplifying the magnetic fluctuations that would allow the expected scattering. However we found stringent conditions on the mean field for that amplification. 

The recent remarkable work by A. Spitkovsky  \cite{AS} shows through PIC simulations, that the Fermi process works at relativistic shocks in a non-magnetized pair plasma. The shock structure is quite different. The magnetic fluctuations are produced by a Weibel instability. We will indicate how weak a mean field must be for the instability to be operative even with a proton-electron plasma. 
We propose also another instability that works with a larger mean field, however still quite weak.

 There is observational evidence in Gamma Ray Bursts, especially for the afterglows, that both the magnetic field has been amplified upstream and that power law distributions of high energy particles has been formed suggesting a Fermi process (see, for instance Li \& Waxman 06 \cite{LW}). Moreover GRBs often explode in  Wolf-Frayet star winds, which are much more magnetized than the interstellar medium (Ud-Doula et al. 08).

\section{Phase space locking by the transverse mean field }

\subsection{How a collisionless shock is set up?}

A collisionless shock is built with the reflection of a fraction of incoming particles at some barrier, generally of electrostatic or magnetic nature, except in the case of a pure electron-positron plasma without any mean magnetic field. It suffices that some downstream obstacle constrains the flow to slow down and thus to compress.
In a proton-electron plasma carrying an oblique magnetic field a barrier of both electrostatic and magnetic nature raises. Because the magnetic field is frozen in most part of the plasma, its transverse component is amplified by the velocity decrease; thus a magnetic barrier is set up.  A fraction of the incoming protons is reflected back. A potential barrier raises also because the electron distribution is close to Boltzman equilibrium and thus $e\Phi \simeq T_e \log {n \over n_0}$ with an electron temperature that grows to a value comparable, but likely different, to that of protons, which reaches $T_p \sim (\Gamma_s-1)m_pc^2$. The potential barrier grows up to a value that allows reflection of a significant part of the incoming protons, which means $e\Delta \Phi \sim (\Gamma_s-1)m_pc^2$ (we assume an ionic population mostly composed of protons). This potential barrier reflects also a fraction of protons, but favors the transmission of electrons that would otherwise be reflected by the magnetic barrier also. The reflection of a fraction of the protons insures the matter flux preservation against the mass density increase downstream. However because the magnetic field is almost transverse, an intense electric field $E=\beta_s B$ energizes these reflected protons such that they eventually cross the barrier. Interactions between the different streams of protons generates a turbulent heating of the proton population, which takes place mostly in the ``foot" region. The foot region extends from the barrier
upstream over a length $\ell_F = r_{L,F} \equiv \Gamma_s V_s/\omega_{ci,F}$, measured in front frame, and measured in upstream frame, it leads to
\begin{equation}
\label{LF}
\ell_{F,u} = r_{L,u}/\Gamma_s^3 = \frac{c}{\omega_{ci} \Gamma_s \sin \theta_B} \ .
\end{equation}
We assume that the field is almost perpendicular in the front frame, but take into account of an angle $\theta_B$ in the co-moving upstream frame such that $\sin \theta_B > 1/\Gamma_s$.
The downstream flow results from the mixing of the flow of first crossing ions (adiabatically slowed down) with the flow of transmitted ions after reflection. All the ingredients of a shock are realized. The three ion beams in the foot interact through the ``modified two stream instability"; this constitutes the main thermalisation process of the ion population that continues with only two beams passed the potential barrier. These anomalous heating processes require an appropriate kinetic description.\\


{\it Ramp, overshoot, foot.}\\

Entropy production in the shock transition comes from two independent anomalous (caused by collisionless effects) heating processes for electrons and ions. In non-relativistic shocks, electrons reach a temperature larger than ions; however we do not know yet whether this is still the case in relativistic shocks. Moreover they experience a supplementary heating with the convection electric field. Whereas the growth of ion temperature develops on scale $\ell_F$, electron temperature grows on the very short scale $\ell_r$; this defines the ``ramp" of the shock. Actually, because of this strong gradient where an intense transverse electric current is concentrated, anomalous heat transfer occurs through the ramp.
This electron heating is described by Ohm's law in the direction of the convection electric field 
(z-direction), namely
\begin{equation}
\label{ }
\beta_xB + E = {\eta c \over 4\pi} {dB \over dx} \, \, {\rm with} \, \, E=\beta_s B_0 \ ;
\end{equation}
it indicates that the variation scale is the relativistic resistive length given by $\ell_r \equiv {\eta c \over 4\pi} = \delta_e {\nu_{eff} \over \omega_{pe}}$ ($\beta_x<0$). This is a very short scale comparable to the electron inertial length $\delta_e$ when the anomalous resistivity is so strong that the effective collision frequency is of order $\omega_{pe}$. This is the scale at which the electron population experiences Joule heating, this is thus the growth scale of three major quantities, namely, the potential, the magnetic field and the electron temperature. Even if the scale $\delta_e$ is estimated with electrons so hot that they have the relativistic mass $\Gamma_s m_p$ (i.e. $\delta_e \sim \delta \equiv ({\Gamma_s m_p c^2 \over 4\pi n_F e^2})^{1/2}$), it remains smaller than the foot length; indeed
\begin{equation}
\label{ }
\frac{\delta}{\ell_F} \ll 1 \, \Leftrightarrow \, {B_F^2 \over 4\pi} \ll n_F \Gamma_s m_p c^2 =  \rho_u\Gamma_s^2 c^2 \ .
\end{equation}
that strong inequality is a natural requirement for a relativistic strong shock. By the way, the condition for the Fermi process to be operative will turn out to be more restrictive for the field intensity, as will be seen further on.

This strong Joule heating results from the very intense cross-field current and, actually, is produced by microturbulence driven by the current itself. 
Probably an anomalous diffusion of electron temperature occurs that smoothes the temperature profile; however it is not yet identified in relativistic shocks.\\

The field profile can be obtained by prescribing a velocity profile decreasing from 1 to 1/3 over a distance much larger than $\ell_r$. The profile displays a ramp at scale $\ell_r$ followed by an ``overshoot" before reaching the asymptotic value $3B_0$. Electrons have to jump this barrier in order to come from downstream to upstream. The formation of a foot over a length of order $\ell_F $ much larger than the resistive length, which characterizes the ramp and the overshoot, is the main feature of collisionless shocks, perpendicular or oblique (see figure(1).) \\

\begin{figure}
  \includegraphics[height=.3\textheight]{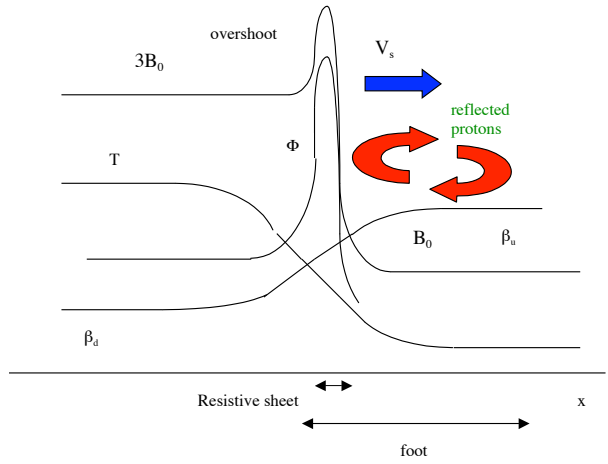}
  \caption{Shock structure: This drawing roughly sketch the profile of the main physical parameters that varies in the foot, the ramp, the overshoot.}
\end{figure}

In the case of an electron-positron plasma, when a magnetic field is considered, no electrostatic barrier raises, only the magnetic barrier occurs. When the mean magnetic field is very negligible, a barrier can raise only through the excitation of waves. This is a crucial aspect of the simulation of A. Spitkovsky \cite{AS} that shows that the Weibel instability can produce particle reflection through the wave growth that exerts a ponderomotive force.

When electron reach energies comparable to $\Gamma_s m_pc^2$, the situation becomes similar to the electron-positron plasma shock.

The strucutre is described with two scales $\delta$ and $\ell_F$ and three small parameters: $\xi_{cr}$,
the fraction of incoming energy density converted into cosmic ray pressure at front, $\xi_B$ the ratio of magnetic energy density over the incoming energy density and $1\over \Gamma_s$.

\subsection{Particle motions}

\subsubsection{Particle motions upstream}

Besides the part of ``thermal" protons that are reflected back on the electrostatic barrier and eventually cross the ramp after having gain an energy of order $\Gamma_s m_pc^2$, some amount of the downstream protons comes back upstream, scatters off the magnetic field and comes back downstream with an energy of order $\Gamma_s^2 m_pc^2$. This is the first cycle of Fermi acceleration, which is always possible for a fraction of the particles and that generates a primary population of high energy cosmic rays. The question is to know whether further Fermi cycles are possible.

Let us first look at the kinematics of cosmic rays upstream in a large scale mean field $\vec B= B \vec e_y$ parallel to the shock plane; it is convenient to make this investigation in the co-moving frame, where the electric field vanishes and where particle orbits are circles characterized by the Larmor radius. The momentum $p_y$ is conserved. With simple geometrical arguments one can derive the gyro-phase variation $\psi$ as a function of the initial angle $\theta_0$ of the momentum in the vertical plane with respect to the flow (see figure 2).

\begin{figure}
  \includegraphics[height=.3\textheight]{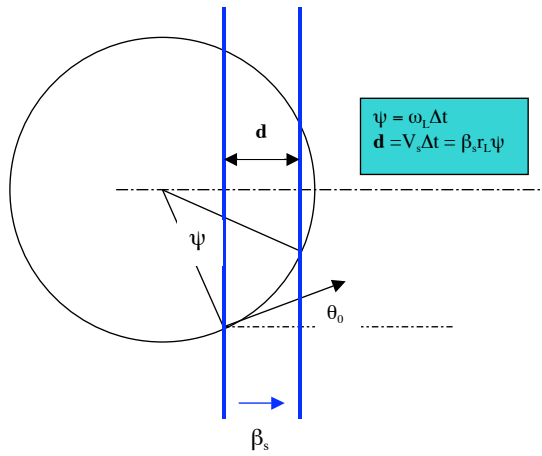}
  \caption{Particle motion upstream}
\end{figure}

One finds
\begin{equation}
\label{ }
\sin (\theta_0+\psi) - \sin \theta_0 = \beta_s \psi \ .
\end{equation}
The angle is necessarily such that $\cos \theta_0 > \beta_s$ and thus $\vert \sin \theta_0 \vert < 1/\Gamma_s$. The small angle approximation leads to
\begin{equation}
\label{ }
{1\over 3} \psi^2 + \theta_0 \psi + \theta_0^2 = {1\over \Gamma_s^2} \ .
\end{equation}
The maximum deviation is obtained for $\theta_0 = 0$ and equals $\pm {\sqrt{3}\over \Gamma_s}$. The protons are deviated upwards, whereas electrons are deviated downwards. One easily checks that no deviation occurs at the maximum entrance angle $\pm 1/\Gamma_s$. The visit time is thus given by $\psi = \omega_L \Delta t \sim 1/\Gamma_s$ and thus $\Delta t \sim t_L/\Gamma_s$. The penetration length of the cosmic rays into the upstream flow is therefore of order $r_L/\Gamma_s$. The distance between the orbit and the shock front is given by $\Delta x = r_L(\sin(\theta_0+\psi) - \sin \theta_0 -\beta_s \psi)$ and reaches a maximum for $\sin (\psi +\theta_0)= {1\over \Gamma_s}$ and thus $\Delta x_{max} \sim {r_L\over \Gamma_s^3}$; this indicates the scale at which magnetic perturbations could scatter the cosmic rays.

This relation allows to determine the energy gain after a Fermi cycle $dud$ as a function of the entrance angle. Using back and forth Lorentz transformations we obtain
\begin{equation}
\label{ }
G(\theta_0) = \frac{1- \beta_r \cos (\theta_0 + \psi)}{1- \beta_r \cos \theta_0} \simeq
\frac{1+ {\Gamma_s^2 \over 2}(\theta_0 + \psi)^2}{1+ {\Gamma_s^2 \over 2}\theta_0^2} \ ,
\end{equation}
where we inserted the relative velocity between the upstream and the downstream flows (actually in the latter formula we inserted a converging velocity $-\beta_r$ and counted $\beta_r$ positively):
$$
\beta_r = (\beta_u-\beta_d)/(1-\beta_u \beta_d) \simeq 1-{1\over \Gamma_s^2} \ .
$$
The angle $\theta_0=0$ produces the maximum gain: $G(0)= {5\over 2}$ and the minimum gain, $G=1$, is obtained for $\theta_0 = \pm 1/\Gamma_s$. The gain monotonically decreases between these two values. The gain averaged over this angle interval is close to the value 2. The large gain of order $\Gamma_s^2$ cannot be reached with particles coming from downstream to upstream; only the first Fermi cycle $udu$ can produce such an extreme gain (see \cite{GAC}, \cite{AGKG}).

\subsubsection{Particle motion downstream, the ``no return" sector}

A similar dynamical system governs downstream motion; $\beta_s$ is replaced by $\beta_d \simeq 1/3$ for a strong relativistic shock. An incoming proton with $\beta_z > 0$ and $\beta_x < \beta_d$ has no possibility to come back to the shock. Only intense magnetic perturbations on short scales (on scale comparable to the Larmor radii of the particles measured in front shock) could produce efficient scattering that would allow further Fermi cycles. For instance if strong turbulence makes the magnetic field to reverse on short scales this would probably make Fermi cycles possible.

A simple geometrical argument can explain the ``no-return" condition (see figure 2). A similar geometrical construction can be done to look at the orbit of an incoming particle that rotates in the co-moving downstream flow whereas the shock is moving away at the speed $c/3$. 

\begin{figure}
  \includegraphics[height=.3\textheight]{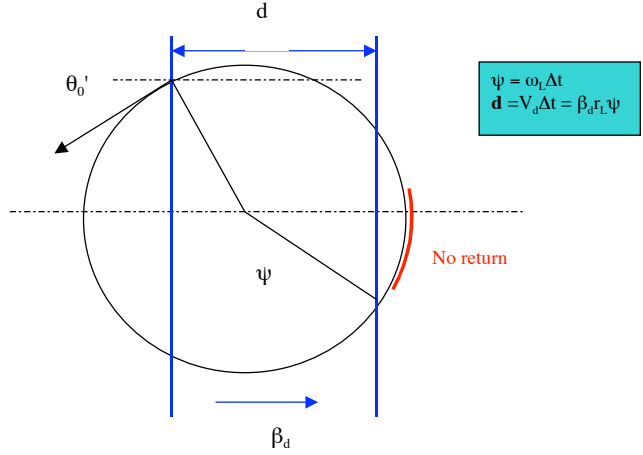}
  \caption{Particle motion downstream}
\end{figure}

One finds the relation
\begin{equation}
\label{ }
\sin (\theta_0+\psi) -\sin \theta_0= {\psi \over 3} \, \, {\rm or} \, \, \beta_d \psi
\end{equation}
with the condition that incoming particles are such that $\cos \theta_0 > {1/3}$. This equation has roots in some interval and no roots in the complementary interval. It can be seen first by considering two simple cases. First consider the case $\theta_0 = 0$; the equation has no root because ${\sin \psi \over \psi} > -{1\over 3}$ for all $\psi$. Now consider the case $\theta_0 = {\pi \over 2}$; the graph shows evidence of a root. 

A particle crosses the shock from downstream to upstream if it has an angle $\theta_d \in (-\theta_*, \theta_*)$, such that $\cos \theta_*= 1/3$ ($\theta_* \simeq 1.231 \, rad$), (a ``pitch" angle $\theta$ corresponds to an angle $\alpha -{\pi \over 2}$ on the Larmor circle) which corresponds to $\alpha = \theta -\pi/2$ between $-\theta_*-\pi/2$ and $\theta_*-\pi/2$ on the Larmor circle ($\alpha_*= -0.349 \, rad$). 
Only some interval of initial angles allows the particle to re-cross the shock $d \rightarrow u$. This interval starts at $\theta_*$ and ends at $\theta_{cr}$ such that $\theta_{cr} + \psi = \theta_*$; this angle is thus the root (other than $\theta_*$) of the equation 
\begin{equation}
\label{ }
\sin \theta - \sin \theta_* = \frac{\theta - \theta_*}{3}
\end{equation}
The root is between $\pi/2$ and $\pi$, with $\theta_{cr} \simeq 1.710\, rad$. The interval $(\theta_*, \theta_{cr})$ defines a sector of no-return. The particles that make the cycle $dud$ come back downstream with an angle slightly larger than $\theta_*$ and thus flow in the sector of no-return. It corresponds to angle $\alpha_{cr}$ between 0 and $\pi/2$ on the Larmor circle, $\alpha_{cr} = 0.140\, rad$.

\subsubsection{Requirement on turbulence for Fermi cycles}

The previous paragraph was devoted to motions in a regular transverse field.
However the behavior we have analyzed with an ordered large scale field also applies for ordinary turbulence, that cascades from large scales to short scales. This is an important point that has been argued in Lemoine, Pelletier \& Revenu \cite{LPR}. In particular the figures in this paper clearly show that the Poincar\'e velocity mapping, in a plane parallel to the shock plane, is similar between regular motions and irregular motions in a Kolmogorov turbulence field and reveals the sector of ``no return". 
This explains also the results by Niemiec et al. (\cite{NO}, \cite{NOP}).
In order that particles undergo several Fermi cycles and form a power law distribution, intense scattering should develop in this flow. Magnetic turbulence should be excited for the Fermi process to be operative, upstream at time scale shorter than $t_L/\Gamma_s^3$ over a precursor of length $r_L/\Gamma_s^3$ and downstream at time scale shorter than $t_L$ over scale shorter than $r_L$, with $\delta B$ at that scale more intense than the large scale field...

Thus it is particularly important to look for the generation of intense short scale turbulence upstream, and preferentially {\it incompressible} turbulence that can reach higher level than compressible one.
The requirement on the relative amplitude $A$ of the magnetic fluctuation can be explored by looking at the variation of the angles $\theta$ and $\phi$ defined by 
$\beta_x \cos \theta$, $\beta_y = \sin \theta \cos \phi$ and $\beta_z = \sin \theta \sin \phi$.
\begin{eqnarray}
< \Delta \phi^2 > & \simeq & {1\over 3}A^2 {\omega_L^2 \over \theta^2} \tau_c \Delta t \\
< \Delta \theta^2 > & \simeq & {2\over 3}A^2 \omega_L^2 \tau_c \Delta t  \ .
\end{eqnarray}
Careful analysis \cite{PLM} shows that the amplitude must be quite high, namely $A > r_L/\Gamma_s \ell_c$, $\ell_c$ being the coherence length of the magnetic fluctuations, where the Larmor radius is estimated in the quadratic averaged field and Fermi acceleration is restricted to limited range of Larmor radii:
\begin{equation}
\label{ }
\Gamma_s \ell_c < r_L < A \Gamma_s \ell_c \ .
\end{equation} 

\section{Instabilities with a mean field}

The expected instability that would scatter the cosmic rays upstream is supposed to do it during the precursor crossing. Two states of the precursor must be considered: one (case A) is the stationary situation where Fermi process is supposed to be developed such that the precursor length is determined by the diffusion length of the most energetic Cosmic Rays; the other (case B) is the starting situation where the cosmic rays of the first Fermi cycle are scattered by the mean field. 

In case A, when particles are scattered off short scale, but intense, magnetic fluctuations, the scattering frequency is 
\begin{equation}
\label{ }
\nu_s \sim c e^2 <\delta B^2> \ell_c/E^2
\end{equation}
and the diffusion length $\ell_d \sim c/\nu_s$ (the diffusion coefficient is $D=c^2/3\nu_s \propto E^2$); but the scattering must develop over a length of order 
$c/\nu_s\Gamma_s^2$ because of the front motion.  For a temporal growth rate $\gamma_{inst}$, the growth factor  is
\begin{equation}
\label{ }
G_{inst} = \frac{\gamma_{inst} c}{\nu_s V_s \Gamma_s^2}
\end{equation}
and the energy of the magnetic fluctuations is amplified by a factor ${\cal A}_{inst} \sim e^{2G_{inst}}$ in the linear regime. The instabilities are efficient if
\begin{equation}
\label{ }
\gamma_{inst} > \omega_{ci} \frac{V_sV_A}{c^2} \frac{\ell_c}{\ell_{mhd}} \frac{<\delta B^2>}{B_0^2}
\frac{\Gamma_s^2}{\gamma^2} \ ,
\end{equation}
where $V_A$ is the Alfv\'en velocity, $\ell_{mhd} \equiv V_A/\omega_{ci}$ is the minimum scale for MHD description.
Even MHD instabilities could develop.

In case B, (see eq.\eqref{LF}) the growth factor
\begin{equation}
\label{ }
G_{inst} = \frac{\gamma_{inst} c}{\omega_{ci} V_s \Gamma_s \sin \theta_B} ;
\end{equation}
and thus the temporal growth rate must be larger than
\begin{equation}
\label{ }
\gamma_{inst} > \omega_{ci} \beta_s \Gamma_s \sin \theta_B \ .
\end{equation}
Note that we keep $\Gamma_s \sin \theta_B > 1$ throughout the further discussions. This last constraint is the most severe.
Any MHD mode, whose temporal growth rate is necessarily smaller than $\omega_{ci}$, cannot be amplified unless the magnetic field would be parallel to the flow. Nor any extraordinary ionic mode that would have a frequency close to $\omega_{ci}$, such a mode with a frequency close to the lower hybrid frequency would be better. Actually if the scattering process develops through any fast enough instability, it enlarges the precursor and makes the instability growth easier.

Of course the electronic modes are fast enough, but their capability to scatter high energy cosmic rays is very unlikely. 

\subsection{At MHD scales}

For the frequently valid condition $\beta_A \Gamma_s \sin \theta_B \ll 1$, the precursor has a length much larger than the minimum scale for MHD description ($\ell_{mhd}/ \ell_{F,u} = \beta_A \Gamma_s \sin \theta_B $). The growth of MHD modes is expected mostly in case A. There is an interesting possibility of Alfv\'en mode amplification due to the quasi Tcherenkov resonance that selects Alfv\'en modes such that $\omega_A = k_{\parallel}V_A = k_xV_s$. The growth rate is 
\begin{equation}
\label{ }
\gamma_{inst} \sim (\beta_s^2 \omega_{p*}^2 \omega_A)^{1/3} \ ,
\end{equation}
where $\omega_{p*} \equiv (4\pi n_{cr}e^2/\gamma_*m_p)^{1/2}$ with $\gamma_* \sim \Gamma_s^2$
(there is a factor $\Gamma_s$ because the energy of the particles in the front frame is of order $\Gamma_s m _p c^2$, and another $\Gamma_s$ because of the frame motion). The cosmic ray density $n_{cr}$ is measured in front frame and is related to the cosmic ray pressure at front frame $P_{cr}$ by 
$P_{cr} = n_{cr} \Gamma_s m_p c^2 = \xi_{cr} \rho_u \Gamma_s^2c^2$, so that $\xi_{cr} = n_{cr}/n_u\Gamma_s$.
The frequency $\omega_A$ is smaller than $\omega_{ci}$ but can be close, especially for the right mode.\\

When the magnetic field is almost parallel, i.e. when $\theta_B < 1/ \Gamma_s$, Bell's instability \cite{BELL} \cite{RI} \cite{ZPV} (see \cite{RE} in relativistic regime) can develop. It is excited by the charge current carried by the cosmic rays in the precursor. Non-resonant waves of wavelength shorter than the Larmor radii are exited without any response of the cosmic ray plasma. Its growth rate in upstream frame is of order $k_cV_A$ with $k_c=4\pi J_{cr}/cB_u$. Unfortunately this field configuration is not generic for relativistic shock. However for a quasi perpendicular field, in the same spirit as Bell's instability, the electric charge carried by the cosmic rays can excite non-resonant compressive modes for which the cosmic rays have no response because the waves are at short scales compared to the Larmor radii also. The cosmic-ray electric charge is almost neutralized by the ambient plasma electric charge in the front frame and a typical wave number is associated to it: $k_* \equiv {n_{cr}eB_0/\rho_u \Gamma^2_s c^2}$ (defined in front frame). The spatial growth rate measured in front frame is (see Pelletier, Lemoine, Marcowith 08 \cite{PLM})
\begin{eqnarray}
\label{ }
\gamma_x & \sim & (k_*^2k_x)^{1/3} \, {\rm for} \, \, k_y \ll (k_*k_x^2)^{1/3} \\
           & \sim & (k_*k_y)^{1/2} \, {\rm for} \, \, k_y \gg (k_*k_x^2)^{1/3} \, 
\end{eqnarray}
(x is along the flow and y along the mean field).

In case A, the growth requirement is likely fulfilled since 
\begin{equation}
\label{ }
k_*{r_L^2 \over \ell_c} = {n_{cr} \over n_u}{1\over \beta_A \Gamma_s} = {\xi_{cr} \over \beta_A} \ ,
\end{equation}
and is larger than 1 for $\beta_A \Gamma_s$ sufficiently small. In case B this is more difficult because 
$k_*\ell_F = {n_{cr} \over n_u}{1\over \Gamma_s} = \xi_{cr}$, which is rather small, but this can be compensated by the wave numbers that can be much larger than $k_*$.

The development of the compressive instability in case A is not expected to solve the scattering issue. It produces more likely heating (see further on for a more detailed discussion) and the magnetic field fluctuations remain moderate because of the limitation of the density depletion. However a nonlinear investigation, through numerical simulations, should be useful to look more deeply in the consequences of the instability.

\subsection{At micro-scales}

When the ambient magnetic field is disregarded, the reflected particles and the fraction of particles that participate to the first Fermi cycle, constitute a relativistic cold beam that pervades the ambient plasma and trigger micro-instabilities. One is the two stream instability, which mostly amplifies the electrostatic Langmuir field through a resonant interaction such that $\omega- k_x V_s = 0$ (Landau-Tcherenkov resonance) and also an electromagnetic component because of the density inhomogeneity. The other is the Weibel instability, with $k_x = 0$ (it is non-resonant or the resonance is somehow reduced to $\omega=0$), which is mostly electromagnetic with a low phase velocity so that the magnetic component of  the wave is dominant. That instability is suitable for developing particle scattering. Its growth rate is of order $\beta_s \omega_{p*}$, where $\omega_{p*} \equiv (4\pi n_{cr}e^2/\gamma_*m_p)^{1/2}$ with $\gamma_* \sim \Gamma_s^2$.

When the ambient magnetic field is considered, the relativistic stream of particles that pervades the upstream plasma is also seen as a cold beam from the upstream comoving frame, with a transverse dispersion of order $1/\Gamma_s$, but with a finite penetration length ($\ell_{F,u}$). The beam is almost non-magnetized and can undergo quasi Tcherenkov resonance, as previously, but with waves of the ambient plasma where the magnetic field plays a role. Depending on the nature of the excited waves, the ambient plasma is more or less magnetized with respect to the wave dynamics. Ambient electrons are easily magnetized, however it turns out that even magnetized, Weibel instability can develop provided that ambient protons are not (see \cite{LPM}). Between the inertial scale, $\delta_e \equiv c/\omega_{pe}$, of electrons and the inertial scale $\delta_i \equiv c/\omega_{pi}$ of the ions, there is an intermediate dynamical range between MHD (at scales larger than $\delta_i$) and electron dynamics (i.e. dynamics for which electron inertia is relevant), where fast enough waves can be excited by the relativistic stream.
In the continuity of right Alfv\'en waves (the left ones are completely absorbed at the ion-cyclotron resonance), for quasi parallel (with respect to the mean field) propagation there are whistler waves, that are electromagnetic waves with a dominant magnetic component. For quasi perpendicular propagation, there are the ionic extraordinary modes, which have frequencies between the ion-cyclotron frequency and the low-hybrid frequency (obtained for large refraction index) and which are mostly electrostatic with a weaker electromagnetic component.

For scattering purpose, the whistler waves are the most interesting in this intermediate range, (they are excited in the foot of collisionless shocks in space plasmas), and for pre-heating purpose, the extraordinary ionic modes are more interesting (they are actually used for additional heating in Tokamak), as in collisionless non-relativistic shocks in solar wind.

\subsection{Whistler modes instability}

In usual conditions, whistler waves, that are right modes, resonate at Landau-synchrotron resonance with electrons only. In this peculiar situation, they resonate at the quasi Tcherenkov resonance, as mentioned above (see \cite{LPM} for details on these developments). Therefore all relativistic particles participate in the interaction and thus experience scattering if the waves are excited by the stream.
Actually a fast instability results from the resonance with the following characteristics:
a frequency $\omega_w = \omega_{ci} k^2\delta_i^2 = \omega_{ce} k^2 \delta_e^2$, $k_x = \omega_w/c \ll k$ and a growth rate
\begin{equation}
\label{ }
\gamma_{inst} \sim (\beta_s^2 \omega_{p*}^2 \omega_w)^{1/3} \ ,
\end{equation}
within some unimportant angular factors.
This instability exists in presence of a mean field and its growth rate looks a little faster than the Weibel growth rate when
\begin{equation}
\label{ }
\frac{B_u^2}{4\pi} (k\delta_i)^4 > \frac{n_{cr} m_pc^2}{\Gamma_s^2} \ .
\end{equation}
However, when the field satisfies this inequality, the Weibel instability is even quenched (see further on) and the instability of whistler modes alone will grow both in case A and B.

\subsection{Instability of extraordinary ionic modes}

At large wave-lengths, the extraordinary ionic waves become the magneto-sonic waves of MHD description. They are excited by the resonant interaction with the beam. The growth rate is a sizable fraction of the mode frequency $\omega_x$, with $\omega_{ci} < \omega_x < \omega_{lh}$ (where the low-hybrid frequency is such that $\omega_{lh}^2 \simeq \omega_{pi}^2 + \omega_{ci}^2$, with the assumption $\omega_{pi} \gg \omega_{ci}$). Whereas a non-relativistic beam selects a resonant mode of high refractive index, which corresponds to a high frequency close to $\omega_{pi}$, a relativistic one selects a resonant mode of refractive index close to 1, which corresponds to a frequency slightly larger than $\omega_{ci}$. Thus whereas hybrid waves are strongly excited in a non-relativistic collisionless shock foot, lower extraordinary modes are moderately excited in a relativistic shock. They can nevertheless develop more intensively in case A and their instability corresponds to the micro-scale  version with resonance of the previous compressive instability. Like for its MHD scale version, mostly  heating is expected from this instability; by heating we mean mixing of the streams in  the foot region.

\section{Obstruction by the mean field}

In this section we analyze the obstruction to the instability growth by the mean field (case B). Let us see first how it prevents Weibel instability.

The condition that the pervading cosmic rays are not magnetized at the time scale the Weibel instability growth is naturally fulfilled since it requires
 \begin{equation}
\label{ }
\frac{B_u^2}{4\pi} \ll n_{cr}m_p\Gamma_s^2c^2 \ .
\end{equation}

A necessary condition for Weibel instability to develope is that the ambient protons are non-magnetized at the time scale of the instability, which requires $\beta_s \omega_{p*} \gg \omega_{Lu}$, namely
\begin{equation}
\label{ }
\frac{B_u^2}{4\pi} \ll \frac{n_{cr}m_pV_s^2}{\Gamma_s^2} \ .
\end{equation}

The most severe requirement is that the instability has time to grow significantly during the precursor crossing, namely:
\begin{equation}
\label{ }
\beta_s \omega_{p*} {r_{L,u} \over \Gamma_s^3 V_s} > 1 \ ,
\end{equation}
which generally implies a very small mean field:
\begin{equation}
\label{WEIN}
\frac{B_u^2}{4\pi} \sin^2 \theta_B \ll \frac{n_{cr}m_pV_s^2}{\Gamma_s^4} \ .
\end{equation}
When this condition is fulfilled, then the condition of non-magnetization of ambient protons is fulfilled.
It has been shown \cite{LPM} that the ambient electrons can be magnetized; this does not quench Weibel instability.

As for the whistler instability the analogous requirement is
\begin{equation}
\label{WHIN}
\frac{B_u^2}{4\pi} \sin^2 \theta_B \ll \frac{n_{cr}m_pV_s^2}{\Gamma_s^4} \frac{k^2\delta_i^2}{\Gamma_s} \ .
\end{equation}
Since $k^2\delta_i^2$ can reach the value of the mass ratio $m_p/m_e$,  for $\Gamma_s < m_p/m_e$ the whistler modes are less inhibited by the mean field than the Weibel waves. Nevertheless even with whistler mode excitation, the intensity of the mean magnetic field needs to be quite low in order that waves have time to grow, except, of course when the shock Lorentz factor is weak.

Short scale magnetic fluctuations can undergo a fast growth in an ultrarelativistic shock precursor only if the ambient mean field is very weak; then the relativistic Fermi process is operative. The criteria is the following. Let $X \equiv {m_e \over m_p} \Gamma_s$ and $Y \equiv {\Gamma_s^4 B_u^2
\sin^2\theta_B / 4\pi n_{cr}m_pV_s^2}$; then  Weibel waves grow if $Y < 1$;  whistler modes grow if $XY <1$ and the previous growth rate is valid for $Y>X^2$ (otherwise there is no resonance); all the instability regimes are sketched in fig.4.

\begin{figure}
  \includegraphics[height=.3\textheight]{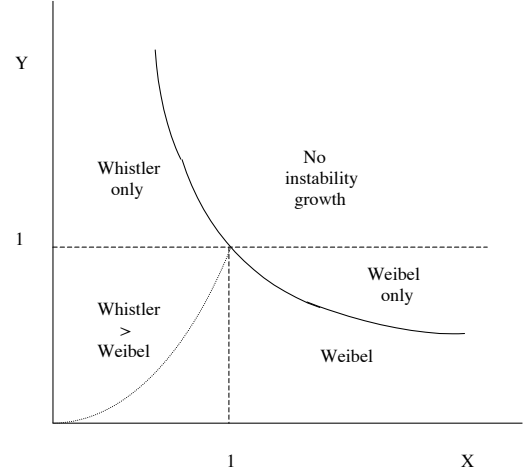}
  \caption{Instability diagram: $X \equiv {m_e \over m_p} \Gamma_s$, $Y \equiv {\Gamma_s^4B_u^2 \sin^2 \theta_B/ 4\pi n_{cr}m_pV_s^2}$. Weibel instability develops for $Y<1$, 
  the whistler instability for $XY<1$ and $Y>X^2$}
\end{figure}

This is only in case A that we can expect a significant excitation of extraordinary ionic modes and magneto-sonic modes. They generate a pre-heating process in the precursor that concerns both electrons and protons. We cannot say at this stage of the investigation whether the incoming plasma is strongly heated in the shock foot. If electrons and protons are heated at  similar relativistic temperature, it changes the instability dynamics. Moreover it would start to slow down significantly the upstream flow.

\subsection{Hot plasma in shock foot}

We do not know yet the details of a relativistic shock front. In the previous section we assumed that the shock front is structured like a non-relativistic front and just extended the non-relativistic results. Since MHD compressive instability and extraordinary ionic modes can be excited, we cannot exclude that the foot be full of relativistically hot protons and electrons of similar temperature ($\bar \gamma m_pc^2$). In that case the plasma response is different, because the intermediate whistler range (and also extraordinary range) disappears; the plasma behaves like a relativistic pair plasma. In that case, only Weibel instability is expected and the limit on the intensity of the ambient magnetic field for the instability growth is such that
\begin{equation}
\label{ }
\frac{B_u^2}{4\pi} \sin^2 \theta_B < n_{cr}m_pc^2 {\bar \gamma^2 \over \Gamma_s^4} \ .
\end{equation}
If this happens, this would  be the ultimate criteria for developing Fermi acceleration at relativistic shocks.

\section{What happens downstream?}

In the downstream plasma, the magnetic fluctuations generated by Weibel instability disappear rapidly because they do not correspond to plasma modes. However whistler waves are transmitted and although they are not excited downstream, their damping is weak. 
When Fermi cycles develop, they create ``inverted" distribution downstream, that should produce a maser effect. Otherwise the intrinsic magnetic fluctuations would decay, only extrinsic MHD turbulence would scatter large Larmor radius particles.

Tangled magnetic field carried by the upstream flow are very compressed downstream and thus opposite polarization field lines come close together. This produce magnetic reconnections in an usual regime where protons and electrons have a similar relativistic mass of order $\Gamma_s m_pc^2$.
Such a regime of reconnection deserves a specific investigation with appropriate numerical simulations. Despite magnetic dissipation, reconnections would probably create a chaotic flow that favors diffusion of particles from downstream to upstream. \\

\section{Discussion}

Because the ambient mean magnetic field has a dominant transverse component at the relativistic shock front, the scattering of particles that are expected to undergo Fermi cycles requires very intense magnetic fluctuations at short scales. Short scales instabilities can be excited by the Fermi cycles, and eventually by the first Fermi cycle. However the mean transverse field again severely limits the efficiency of the instabilities by reducing drastically the length of the precursor. We found two regimes of efficient excitation of magnetic fluctuations suitable for scattering: one dominated by Weibel instability, the other by the resonant instability of the whistler modes. If particles are not too much pre-heated in the precursor, the latter is less inhibited by the mean field than the former. However it is also severely limited and no Fermi acceleration can occur, unless the mean field intensity is such that:
\begin{equation}
\label{ }
\frac{B_u^2}{4\pi}  \ll {m_p\over m_e} \frac{n_{cr}m_pV_s^2}{\Gamma_s^5 \sin^2 \theta_B} \ .
\end{equation}
Now if electrons and protons are heated to relativistic temperature in the shock foot, only Weibel instability can make Fermi process to work provided that the mean field is so weak that
\begin{equation}
\label{ }
\frac{B_u^2}{4\pi}  < {n_{cr}m_pc^2 \over \sin^2 \theta_B} {\bar \gamma^2 \over \Gamma_s^4} \ .
\end{equation}
Indeed we found instabilities that can heat the precursor.
As can be seen, the angle of the magnetic field with respect to the shock normal is a sensitive parameter. It turns out that the perpendicular collisionless shocks are unsteady in non-relativistic regime. They are probably unsteady in relativistic regime also. This is related to the development of whistler waves (\cite{LKB} \cite{SCH}). The shock reformation occurs at a time scale comparable to the particle crossing time in the foot. Such investigation for a relativistic shock would be important and would deserve heavy numerical simulations.

Regarding the generation of UHECRs by relativistic shocks, the conclusions of  this study is fairly pessimistic, because even if the Fermi cycles work with intense short scale magnetic fluctuations, the scattering time becomes longer and longer with the square of the particle energy. The Hillas criteria is no longer relevant; it should be replaced by $E_{max} = \Gamma ec \bar B (\ell_c R)^{1/2}$, where $\bar B$ is the quadratic average of the fluctuating magnetic field at the short scale $\ell_c$. This new criteria is of no practical use, however it indicates that the very short scale $\ell_c$ makes the UHECR generation hopeless through this process.

These topics suggest the development of heavy 3D PIC simulations for the nonlinear stage, as already started successfully (\cite{HED}, \cite{DIEC}, \cite{AJ}). 

This talk reveals the interest of developing communications between three disciplines: High Energy Astrophysics, Particle Physics and Space plasma physics.


\begin{theacknowledgments}
  One of us, G.P., acknowledges fruitful discussions on these issues, before the symposium with Heinz Všlk, and during the symposium with A. Bell, L. Drury, J. Kirk, B. Reville, J. Arons, Y. Liubarsky, M. Ostrowski, J. Niemiec.
  \end{theacknowledgments}

\end{document}